\documentclass[twocolumn,showpacs,preprintnumbers,amsmath,amssymb]{revtex4}
\usepackage{graphicx}
\usepackage{dcolumn}
\usepackage{bm}
\usepackage[tight]{subfigure}
\usepackage{amsmath}
\usepackage{verbatim}
\usepackage{color}

\begin{document}

\title{Interaction-induced negative differential resistance in asymmetric molecular junctions}

\author{M. Leijnse$^{(1)}$}
\author{W. Sun$^{(1,2)}$}
\author{M. Br{\o}ndsted Nielsen$^{(3)}$}
\author{P. Hedeg{\aa}rd$^{(1)}$}
\author{K. Flensberg$^{(1)}$}
\affiliation{
  (1) Nano-Science Center, Niels Bohr Institute,
      University of Copenhagen,
      Universitetsparken 5, 
      DK-2100~Copenhagen \O, Denmark\\
  (2) BNLMS, SKLSCSUS, College of Chemistry and Molecular Engineering, 
      Peking University, Beijing 100871, China\\
  (3) Department of Chemistry, 
      University of Copenhagen, 
      Universitetsparken 5,  
      DK-2100~Copenhagen \O, Denmark\\
}
\begin{abstract}
Combining insights from quantum chemistry calculations with master equations, 
we discuss a mechanism for negative differential resistance (NDR) in molecular junctions, 
operated in the regime of weak tunnel coupling.
The NDR originates from an interplay of orbital spatial asymmetry and strong electron--electron interaction, which causes
the molecule to become trapped in a non-conducting state above a voltage threshold. 
We show how the desired asymmetry can be selectively introduced in individual orbitals in e.g., OPE-type molecules by 
functionalization with a suitable side group, which is in linear conjugation to one end of the molecule and  
cross-conjugated to the other end.
\end{abstract}
\maketitle

\section{Introduction}
The field of single-molecule electronics has been expanding rapidly during 
recent years, as techniques to electrically contact and control single molecules in a
transport junction have improved~\cite{Reed97, Stipe98, Park99elmig, Kubatkin03, ONeill05}. 
By studying the electric current, $I$, through the molecule as function of the applied  
bias voltage, $V$, spectroscopic information can be extracted~\cite{Stipe98}. In three-terminal setups 
a gate voltage, $V_g$, can be used to control the electrostatic potential on the molecule, 
allowing for detailed spectroscopy~\cite{Park99elmig, Kubatkin03, ONeill05}.

A long standing problem in molecular electronics is how to
design molecules with reproducible and distinctive transport
characteristics, e.g., diode-like rectification as in the original proposal
of Aviram and Ratner~\cite{Aviram74}. 
Another interesting non-linear transport characteristic is negative differential resistance (NDR), i.e., a current which 
\emph{decreases} with \emph{increasing} voltage. NDR is known to occur in resonant tunneling diodes (RTDs) and has in this 
context been suggested for a large number of device applications, such as high-frequency oscillators~\cite{Brown91}, 
logic circuits~\cite{Mathews99}, and analog-to-digital converters~\cite{Broekaert98}. 
Also molecular junctions have been found to exhibit NDR mediated by a mechanism similar to the working principle of RTDs:
a voltage-induced alignment of narrow bands in the electrode and molecule is followed by a mis-alignment
at higher voltages, causing the current level to drop~\cite{Bedrossian89}.
Such mechanisms can, however, not explain the huge NDR effect observed in different derivatives of 
oligo(phenyleneethynylene) (OPE), first observed by Chen et. al.~\cite{Chen99}. 
Here explanations have instead involved either voltage-induced conformational 
changes~\cite{Seminario00, Khondaker04}, or charge-injection~\cite{Chen99, Tour01, Seminario01}, both mechanisms forcing the 
molecule into a less conductive state above a certain voltage threshold. 
Clearly, understanding the origin of NDR in e.g., such OPE molecules, being distinctly different from that of standard RTDs, 
is of great importance. The situation is, however, complicated by inconsistencies in the experimental findings, and
while some experiments observe NDR in individual OPE-type molecules~\cite{Ramachandran02}, others seem to indicate that 
NDR is only observed when large assemblies of such molecules are measured~\cite{Mujamdar06}.

In the above works the tunnel coupling between molecule and electrodes was rather strong. 
In gated (three-terminal) single-molecule junctions~\cite{Park99elmig, Kubatkin03, ONeill05}, 
the tunnel coupling is often much smaller, on the same scale as the thermal energy $k_B T$.  
NDR in this parameter regime has been theoretically discussed in various different types of molecules and can originate from e.g,
Pauli spin-blockade~\cite{Muralidharan07} (a well-known tool for spin-detection in semi-conducting quantum dot 
systems~\cite{Ono02, Hanson07rev}), vibrational blockade~\cite{Wegewijs05, Schultz07a}, or combinations of spin and vibrational 
effects~\cite{Reckermann09a}. Additionally, interference between orbitally degenerate states has been predicted to give 
rise to NDR~\cite{Begemann08} (analogous to quantum dots with ferromagnetic electrodes~\cite{Braun04set}). 
There are also several experimental observations of NDR in molecular junctions operated 
in this regime, see e.g., Refs.~\cite{Heersche06, Osorio07b}.

\begin{figure}[t!]
  \includegraphics[height=0.4\linewidth]{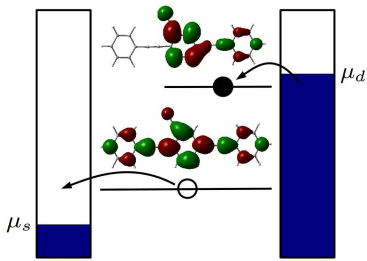}
  \caption{
    \label{fig:1}
    Illustration of the NDR mechanism.     
    Shown are both the molecular structure, with frontier LUMO (bottom) and LUMO+1 (top) orbitals, 
    and a level-diagram, indicating the occupation of these orbitals. 
    The HOMO (not shown) is filled with two electrons and transport involves the neutral and singly reduced molecular charge states.
    Rectangles indicate the source and drain conduction bands with chemical potentials $\mu_s$ and $\mu_d$.  
    Below a (negative) threshold voltage, -$|V_\text{th}|$, an electron tunneling out to the drain from the 
    LUMO can be followed by one tunneling into 
    the LUMO + 1 from the source. Due to the orbital asymmetry,
    the electron in the LUMO+1 cannot leak out into the source and, due to the strong Coulomb interaction,
    it blocks additional charge transfer to the molecule, leading to a dramatically reduced current 
    for $V < -|V_\text{th}|$.
    }
\end{figure}

In this work, we show that intrinsically asymmetric molecular junctions, such as OPE functionalized with appropriately 
chosen side groups, give rise to $I(V)$ curves showing large NDR effects for low temperatures and weak tunnel coupling. 
Above a voltage threshold, a charge carrier (electron or hole) can tunnel onto the molecule and get trapped 
in a molecular orbital (MO) which has a significant overlap with the lead states of one contact only. The Coulomb interaction 
prevents additional charge transfer to the molecule, thereby blocking transport and leading to NDR, see Fig.~\ref{fig:1}. 
Thus, our proposed mechanism arises \emph{because of} strong Coulomb interaction between electrons on the molecule. 
Despite screening by conduction electrons in the leads, the Coulomb energy associated with charging molecules similar to those 
considered here is indeed large (hundreds of meV per added electron or hole)~\cite{Kubatkin03, Kaasbjerg08}. 
Nonetheless, electron--electron interaction is often neglected, or treated on a mean-field level, in popular approaches to 
transport theory of molecular junctions, which would therefore fail to correctly describe the effect discussed here.
We instead use a master equation (ME) approach and calculate the nonequilibrium occupations of the molecular \emph{many-body} 
(i.e., including electron--electron interaction) energy-eigenstates, $| N \alpha \rangle$, where $N$ is the number of valence 
electrons on the molecule and $\alpha$ 
labels the different $N$-electron eigenstates. From these occupations, the current can be calculated in the non-linear regime. 
Thus, both Coulomb interactions and the non-equilibrium condition are included non-perturbatively. 
With knowledge of the mechanism underlying the NDR, 
we use density functional theory (DFT) to investigate different OPE-type derivatives and find molecular structures where the 
effect is maximally pronounced.
An ab initio approach to transport based fully on DFT is inadequate for our purposes because of the 
mean-field treatment of interactions. 
However, DFT can be used to predict the spatial structure of the (single particle) MOs, $\psi_i$.
DFT can thus provide input parameters to a many-body transport theory such as ME.
In Appendix~\ref{sec:transporttheory}, we provide the details of the ME approach,  
as well as the link to the DFT calculations, i.e., the connection between $\psi_i$ and $| N \alpha \rangle$.
We also show how DFT can be used to estimate the the ratio 
$\gamma^r_i / \gamma^r_j$, where $\gamma_i^r$ is the 
hybridization strength between MO $\psi_i$ and the continuum of electron states in electrode $r$, and we give the
relation between $\gamma^r_i$ and the many-body tunnel couplings, $\Gamma_{N \alpha, N' \alpha'}^r$, 
determining the time-scale associated with a transition between states 
$| N \alpha \rangle$ and $| N' \alpha' \rangle$ as a result of electron tunneling.

\section{Orbitals and stability diagrams}
\begin{figure}[t!]
  \includegraphics[height=0.7\linewidth]{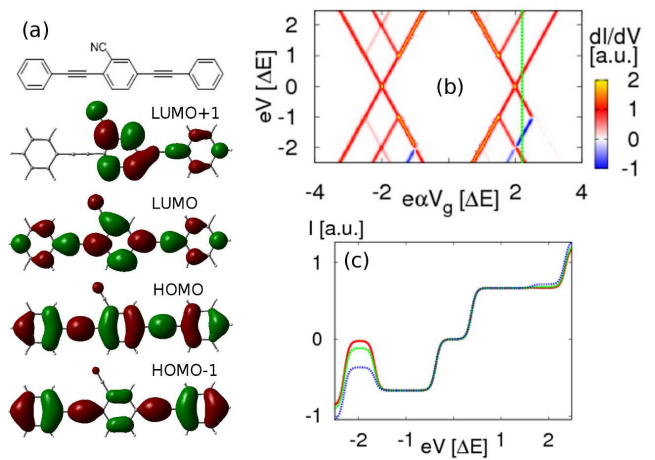}
  \caption{
    \label{fig:2} 
    (a) Molecular structure and frontier orbitals for OPE with a CN side group. (b) Stability diagram, i.e., 
    $dI/dV$ as function of $V_g$ and $V$. The DFT calculations predict
    a very asymmetric LUMO+1 and we set all electrode hybridizations equal to $\gamma$, except 
    $\gamma^s_{\text{LUMO+1}} = \gamma / 10$.
    (c) Red solid curve shows source--drain electron current, $I(V)$, along the dashed green line in (b). 
    Green dashed and blue dotted curves 
    show the same, but for a larger coupling of the LUMO+1 to the drain, $\gamma / 4$ and $\gamma / 2$ respectively.}
\end{figure}

\begin{figure}[t!]
  \includegraphics[height=0.75\linewidth]{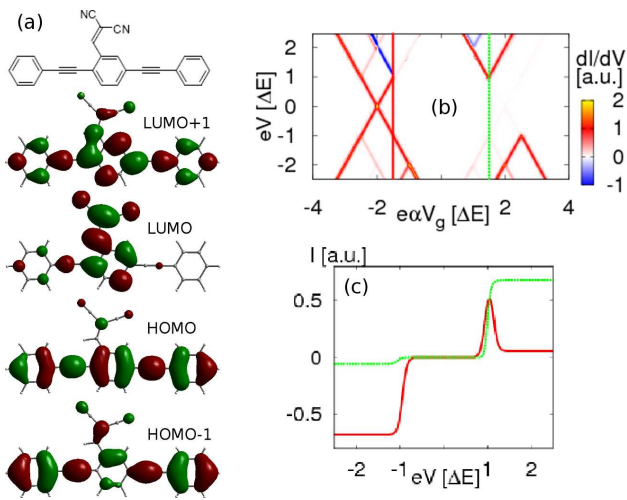}
  \caption{
    \label{fig:3} (a) Molecular structure and frontier orbitals for OPE with a dicyanoethylene side group. 
    (b) Stability diagram, calculated using all lead hybridizations equal to $\gamma$, except $\gamma^d_{\text{LUMO}} = \gamma / 10$.     
    (c) Red solid and green dashed curves show $I(V)$ along the corresponding lines in (b).} 
\end{figure}

\begin{figure}[t!]
  \includegraphics[height=0.5\linewidth]{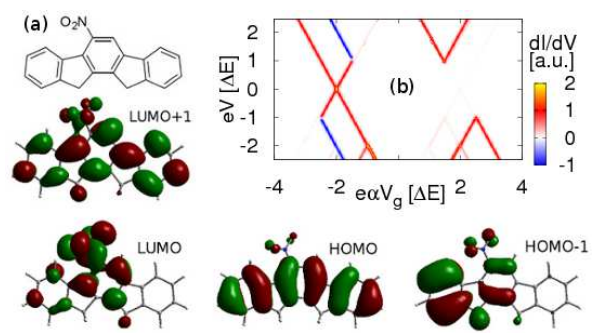}
  \caption{
    \label{fig:4} (a) Molecular structure and frontier orbitals for a rigid molecule, 
    where the benzene rings are 
    fused to five-membered rings, functionalized with a NO$_2$ group.
    (b) Stability diagram, calculated using all lead hybridizations equal to $\gamma$, except 
    $\gamma^d_{\text{LUMO}} = \gamma^d_{\text{HOMO-1}} = \gamma / 10$.} 
\end{figure}

To identify molecules which are expected to show strong NDR effect, we have performed DFT calculations on a large number of 
OPE-type (and similar) molecules, three of which are discussed below. By choosing different side groups, asymmetry can be selectively introduced 
into different orbitals and thereby give rise to blocking states and NDR at different gate and bias voltages. 
Electron-accepting substituent groups are expected to influence most strongly the LUMO and LUMO+1, while electron-donating 
substituent groups are expected to influence most strongly the HOMO and HOMO-1. 
MOs were calculated at the B3LYP/6-311G(2d,p) level using the Gaussian 03 Program package. Visualizations 
of the orbitals for a selected group of interesting molecules are shown in Figs.~\ref{fig:2}--\ref{fig:4}. 

Along with the orbital structure we also show the result of our ME calculations. 
Our purpose here is not to make precise predictions for a specific molecule, but rather to illustrate the general connection
between NDR and the asymmetry of given orbitals.
Rather than using the orbital energies calculated with DFT (see Appendix~\ref{sec:transporttheory}), 
we therefore everywhere assumed energy-equidistant orbitals, 
HOMO-1, HOMO, LUMO and LUMO+1, with energy separation $\Delta E$. All higher/lower orbitals are assumed empty/filled and are not 
included in the calculations, and we focus on the singly oxidized, neutral, and singly reduced charge states only
(for notational simplicity we always refer to the orbitals by their names in the neutral molecule). 
The number of valence electrons included in the ME calculations is thus
$N = 3,4,5$, and the many-body eigenstates are constructed by distributing those electrons over the four orbitals as discussed in 
Appendix~\ref{sec:transporttheory}. 
We assume the charging energies $U_{i j}$ between electrons in MO $\psi_i$ and $\psi_j$ all to be equal, $U_{ij} = 3 \Delta E$. 
In all ME calculations we used the thermal 
energy $k_B T = \Delta E / 50$. Realistic values might be $\Delta E \sim 100$~meV, which would mean $T \sim 20$~K. 
Typically, temperatures $< 4$~K can be reached in experimental setups, but we use a larger value to demonstrate the stability 
of the NDR against thermally activated processes. 
Furthermore, in the regime of intermediate tunnel coupling, $k_B T \lesssim \Gamma$, tunnel broadening (which is not included in the ME) 
would broaden the conductance features and reduce the NDR in a similar way as a larger temperature.
Finally, we simply take the electrode-hybridizations of the MOs to reflect the observed orbital asymmetry, 
i.e., for asymmetric orbitals they are taken much smaller at the lead attached to the side of the molecule with a small orbital weight.
The tunnel couplings of the many-body states are then calculated from Eq.~(\ref{eq:gammadef}) in Appendix~\ref{sec:transporttheory}.

The results of the ME calculations are conveniently presented as so-called stability diagrams, i.e., as differential 
conductance, $dI/dV$, plotted as function of gate and bias voltage, see e.g., Fig.~\ref{fig:2}(b).
Zero gate voltage has been chosen to correspond to the electrode Fermi levels 
(at zero bias) being halfway between the HOMO and LUMO of the neutral molecule (in reality this alignment depends not only 
on the molecule, but also on the details of the electrodes and electric circuit). 
The left/right crossing point at zero bias corresponds to 
oxidation/reduction of the molecule and the distance between those points is the difference between the ionization energy 
and the electron affinity. The linear (small bias) conductance thus give similar information as can be obtained from voltammetry. 
The scaling factor $\alpha$ (gate coupling) depends on the capacitances of the tunnel junctions~\cite{Hanson07rev}.
The gate-dependent lines at finite bias correspond to the energy condition that an (additional) tunnel process 
becomes possible. This happens when a \emph{molecular} chemical potential, 
$\mu_{N \alpha, (N-1)\alpha'} = E_{N \alpha} - E_{(N-1) \alpha'}$, where $E_{N \alpha}$ is the energy of the state 
$| N \alpha \rangle$,
comes into the bias window. Usually, this leads to a stepwise increase of the current, giving a peak in $dI / dV$. 
From the voltage positions of these lines one can thus extract the many-body excitation energies and
a three-terminal molecular junction therefore works as a spectroscopic tool.
The slopes of the lines depend on the capacitances of
the tunnel junctions, which can therefore be extracted in an experimental situation and used to calculate the gate 
coupling $\alpha$~\cite{Hanson07rev}. For simplicity we here assume the capacitances associated with the source and 
drain contacts to be equal,
and equal for all states, and furthermore apply the bias voltage in a symmetric fashion, $\mu_s = eV / 2$, $\mu_d = -eV / 2$, 
where $-e$ is the electron charge, leading to a uniform slope of all $d I / d V$ lines.

In the first example, shown in Fig.~\ref{fig:2}(a), OPE has been functionalized by a cyano (CN) group.
The weights of all orbitals are almost left-right symmetric, with the exception of the LUMO+1, which almost completely 
lacks weight on one of the end benzene rings,
namely the end which is in linear  
conjugation to the cyano group (\emph{ortho} configuration). The other end, with  
all the LUMO+1 weight, is cross-conjugated to the cyano group (\emph{meta}
configuration).  This dependence of the LUMO and LUMO+1 on the conjugation  
pathway to the substituent group is also evident from other structures  
described below. In general, the LUMO always has weight at the  
"\emph{ortho}-end", while the LUMO+1 always has weight on the "\emph{meta}-end" of the  
molecule. 
Figure~\ref{fig:2}(b) shows the stability diagram obtained from the model described above. For positive gate and 
negative bias voltage, there is a clear negative (blue) $d I / d V$ line, i.e, NDR. 
The red solid line in Fig.~\ref{fig:2}(c) shows the current as function of bias voltage along the green dashed line in (b). 
At this value of $V_g$, for very small $|V|$, the molecule is reduced and the current suppressed since interaction 
effects (Coulomb blockade) and the discrete nature of the molecular orbitals prevent charging/discharging of the 
molecule. In the regime of weak tunnel coupling discussed here, the molecular orbitals are very narrow in energy and
transport including only virtual charging of the molecule is suppressed. When $|V|$ becomes larger, the molecule can 
fluctuate between the neutral and reduced charge states as electrons tunnel into and out of the LUMO, and a finite 
current starts to flow. For even more negative $V$, the current then goes down (NDR) as 
sketched and described in Fig.~\ref{fig:1}: when the molecule can be reduced by an electron tunneling from the drain 
into the LUMO+1, rather than LUMO, this electron is "trapped" since it cannot tunnel out into the source
due to the orbital asymmetry; additionally, it blocks transport through the other orbitals, since the strong Coulomb repulsion
prevents more electrons from tunneling onto the molecule. The blockade is lifted and the current recovers when the voltage 
allows an electron to tunnel out to the source from the HOMO instead.
For $V>0$ there is no NDR effect and the LUMO+1 is simply not visible in the transport spectrum.
The current in the blockaded region recovers if the LUMO+1 hybridizes more strongly with the drain electron states, 
as is evidenced by the  green dashed and blue dotted curves in Fig.~\ref{fig:2}(c). 
Note that the three curves are almost identical at positive bias. 

Next, in Fig.~\ref{fig:3}(a), we show OPE functionalized by a dicyanoethylene group, which leads to a strongly asymmetric LUMO.
Analogous to the above case of an asymmetric LUMO+1, this gives rise to a pronounced NDR, see Fig.~\ref{fig:3}(b). 
In this case, however, the blocking state is an excited state of the \emph{neutral} molecule, where an electron has 
been promoted from the HOMO to the LUMO, and therefore the NDR is seen instead at negative gate voltages 
(that it occurs at positive bias is simply because there is a small coupling of the LUMO to the drain, rather than source; 
in an experimental situation this is not easily controllable since it depends on how the molecule binds inside the junction).
The red solid curve in Fig.~\ref{fig:3}(c) shows the current along the red solid line in (b).
There is no strong NDR at positive gate voltage. Instead, the low-bias current is strongly suppressed,
the reason being that the asymmetric LUMO is occupied in the \emph{ground state} of the reduced molecule, which therefore acts as a 
blocking state. Here the $I(V)$ characteristic instead look similar to a regular diode, 
being strongly asymmetric between positive and negative bias, see the green dashed curve in Fig.~\ref{fig:3}(c) which shows 
the current along the green dashed line in (b). Interestingly, using the gate voltage, one can control whether the positive or 
negative bias direction gives rise to a large current.

Asymmetric HOMO or HOMO-1 orbitals would give similar stability diagrams as above, but with NDR due to blocking hole, rather than 
electron, states. 

In OPE-type molecules, the central ring can rotate around the triple bonds and the $\sim 90^\text{o}$
rotated species have only a slightly higher ground state energy compared to the planar ones. 
Ref.~\cite{Seminario00}, which studied molecules similar (although not identical) to those discussed above, concluded
that such rotations can change the spatial structure of the orbitals and a conformational change can switch the molecule 
between conducting and non-conducting states. 
A similar explanation was suggested in the experimental work of Ref.~\cite{Khondaker04}. 
However, it is not at all clear that such rotations can take place once 
the molecule resides inside the junction, possibly "lying" on top of the back gate. We emphasize that all our considerations
are for the lowest energy geometry, which is (close to) planar, and the predicted NDR effect \emph{does not rely on conformational 
changes}. 
Indeed, we are able to create a completely rigid molecule, in which three  
benzene rings are "locked" in a planar conformation by fusing them to two  
five-membered rings (corresponding to two fluorene sub-structures). This  
molecule shows spatial orbital asymmetry when the central ring is  
functionalized with a NO$_2$ group (Fig. 4a).
As in Fig.~\ref{fig:3}(b), the asymmetric
LUMO suppresses transport close to the right crossing point and induces NDR at positive bias close to the left one, see Fig.~\ref{fig:4}(b).
Additionally, the asymmetric HOMO-1 gives rise to NDR at negative bias, as tunneling by a hole into this orbital blocks 
hole-transport through the molecule. 

\section{Conditions for NDR}
The type of NDR effect discussed above relies on a meta-stable (blocking) state being occupied.
Therefore, if this state can relax by other means than tunneling, e.g., by photon emission, 
the degree of blocking and therefore the magnitude of the NDR will be diminished. The life-time 
should here be compared with the typical time between electron tunneling events in the 
\emph{unblocked} regime, i.e., with $e/I$. Even for weakly coupled molecules, this is a rather
short time, e.g., a current $I \sim$~nA, corresponds to a time-scale $\sim 10^{-10}$~s, which
should indeed exceed the lifetime of most excited states. 
That excited molecular states are indeed stable on such time-scales was also the conclusion
of the experimental work in Ref.~\cite{Osorio07b}.

Our conclusions concerning the orbital asymmetry were based on the chemistry of the isolated molecule, 
i.e., neglecting the influence of the bonds to the electrodes, the details of which depend on the 
(unknown and uncontrollable) exact bonding geometry.  
However, we here consider the weak coupling regime, meaning that the molecule--electrode interactions are due to
physisorption, rather than chemisorption, and the resulting perturbations of the molecular 
orbital structure should therefore be small. 
Furthermore, as is seen in 
Fig.~\ref{fig:2}(c), it is sufficient for NDR if one coupling amplitude is twice as large as the other,
so the effect should persist even if the intrinsic asymmetry is partially destroyed by the binding to 
the electrodes.

In addition, the proposed mechanism relies on the ends of the molecular chain binding to the electrodes, 
rather than some other configuration.
This can at present not be experimentally controlled and therefore perhaps 
only a subset of all junctions incorporating such molecules actually show NDR. 
It may be advantageous to functionalize the molecules with appropriate end-groups.
If the electrodes are made of gold, one might e.g., use thiol end-groups, connected to the molecular 
chain through saturated carbon atoms, as in a methylene spacer, to keep the effective molecule--electrode
coupling small~\cite{Poulsen09}.

\section{Suggestion for Ab initio calculations}
Finally, we suggest the possibility to investigate transport through the molecules studied here using a 
completely ab initio approach. 
Due to the intrinsic many-body nature of the predicted NDR effect, the standard ab initio description 
of transport, based on combining a mean-field description of the molecule (e.g., using DFT) with 
non-equilibrium Greens functions (NEGFs), cannot be expected to yield correct results. Nonetheless, 
it might be of interest to undertake such calculations for comparison as they may well capture the 
\emph{existence of} NDR. 
Within a mean-field theory, the Coulomb interaction will increase the energy, $E_j$, of the 
conducting orbital, $\psi_j$, once the blocking orbital, $\psi_i$, enters into the bias window: 
$E_j \propto U_{i j} \langle n_i \rangle$, where $U_{i j}$ is the Coulomb interaction between electrons in 
orbitals $\psi_i$ and $\psi_j$ [cf., Eq.~(\ref{eq:energy}) in Appendix~\ref{sec:transporttheory}], and $\langle n_i \rangle$ is 
the average occupation of orbital $\psi_i$. 
This will reduce the current carried by orbital $\psi_j$ and might lead to NDR. Important here is that 
the \emph{non-equilibrium} aspects of transport are treated correctly, i.e., that the transmission 
function, $T(E, V)$, is calculated for finite $V$ based on the non-equilibrium charge distribution, 
as is done e.g., in the implementations discussed in Refs.~\cite{Brandbyge02, Zahid05}. 
Furthermore, the correct description of transport through the types of molecules discussed here, and their intrinsic NDR effect, 
would provide an interesting testing ground for newly developed ab initio approaches which attempt to go 
beyond a mean-field description, see e.g., Refs.~\cite{Yeganeh09, Mirjani10, Thygesen08}.

\section{Summary and conclusions}
We have shown that an interplay of orbital spatial asymmetry and strong electron--electron interaction
can lead to NDR in molecular junctions.
Using DFT we showed that functionalization of OPE-type molecules
with suitable side groups introduces the desired asymmetry in selected individual orbitals.
Using the chemical insights as a starting point, a master equation approach was used to predict the resulting 
three-terminal transport signal.

Although the purpose of this paper was to investigate the transport behavior in the regime of weak tunnel coupling,
we mention the possibility that our results may still explain some of the NDR effects seen in similar molecules, 
e.g., in Refs.~\cite{Chen99, Tour01, Ramachandran02}, despite the larger tunnel coupling in those experiments. However, we emphasize that
NDR has also been seen in OPE-type molecules without orbital asymmetry. Experimental investigation of the type of 
molecules consider here in three-terminal junctions would be very interesting, as the increased level of control achieved 
with the gate electrode would allow more careful comparison to our predictions, and additionally provide the possibility
of tuning the device to the regime of strong NDR.

\section*{Acknowledgements}
We thank Stephan Sauer and Gemma Solomon for helpful discussions. The research leading  
to these results has received funding from the European Community’s  
Seventh Framework Programme (FP7/2007-2013) under the grant agreement  
"SINGLE" no 213609. \\

\appendix
\section{Master equations for the molecular junction}\label{sec:transporttheory}
We consider a molecule which is coupled to a source ($s$) and a drain ($d$) electrode. 
Neglecting at first the molecule--electrode hybridization, the molecule is characterized by a set of \emph{many-body} 
energy eigenstates $| N \alpha \rangle$, 
where $N$ is the number of valence electrons on the molecule (the number of electrons we explicitly include in the calculations) 
and $\alpha $ labels the different $N$-electron eigenstates ($\alpha$ here labels also the spin degrees of freedom). 
We let $\{ \psi_i \}$ be a single particle orbital basis for the molecule and define operators $c_{i \sigma}^\dagger$ 
which create an electron with spin-projection $\sigma$ in orbital $\psi_{i}$ 
(if needed, one could also include an index $\sigma$ on the orbitals, allowing e.g., for different orbitals for $\alpha$ and $\beta$ spin in 
spin unrestricted calculations). 
Furthermore, electrons in orbital $\psi_{i}$ hybridize with the continuum of electron states in 
electrode $r = s,d$ with amplitude $\gamma^r_{i \sigma}$.
The tunnel rate between molecular eigenstates $| N \alpha \rangle$ and $| N' \alpha' \rangle$ 
is then defined as  
\begin{equation}\label{eq:gammadef}
\Gamma _{N \alpha, N' \alpha'}^{r \sigma} = \frac{2 \pi}{\hbar} d_r \left\vert \sum_{i} \gamma_{i \sigma}^{r}
\langle N \alpha |\left( c_{i \sigma}^{\dagger} + c_{i \sigma} \right) |N' \alpha' \rangle \right\vert ^{2},
\end{equation}
where only the first (second) term gives a non-zero contribution if $N' = N-1$ ($N' = N+1$).
Here $d_r$ is the density of states in electrode $r$, which we assume to be energy-independent (an energy dependence 
is not needed for the NDR mechanism discussed here and would change the results only quantitatively).

We consider the regime of weak molecule--electrode tunnel coupling. Here "weak coupling" means that
all $\hbar \Gamma$ are small compared to the energy separation of the states involved and, in addition, smaller than 
the thermal energy scale $k_B T$. 
In this regime, stationary state (time-independent) transport can be described by leading order perturbation theory in $\Gamma$, 
yielding rate (or master) equations~\cite{Bruus04book}, which are solved to obtain the occupation probabilities $P_{N\alpha }$
of the molecular states:
\begin{eqnarray}
\label{eq:rateeq}
	0 &=& \sum_{N' \alpha'} \left( W_{N \alpha, N' \alpha'} P_{N' \alpha'}  - 
			 W_{N' \alpha', N \alpha} P_{N \alpha} \right), \\
\label{eq:probnorm}
	1 &=& \sum_{N \alpha} P_{N \alpha},
\end{eqnarray}
where Eq.~(\ref{eq:probnorm}) enforces probability normalization. Equation~(\ref{eq:rateeq}) simply states that the occupation 
of a state is determined by the balance of in-going and out-going processes (the sum of all tunnel processes going 
into state $|N \alpha \rangle$, weighted by the occupation of the corresponding initial state $|N' \alpha' \rangle$, minus all tunnel processes going 
out of state $|N \alpha \rangle$).
In leading order the rate matrix $\mathbf{W} \propto \Gamma$ describes only tunnel processes involving a single electron. It thus has 
non-zero elements only when $N$ and $N'$ differ by exactly one, and those elements are given by 
\begin{eqnarray}
\label{eq:intunnel}
	W_{N \alpha, (N-1) \alpha'} &=& \sum_r \Gamma_{N \alpha, (N-1) \alpha'}^{r \sigma} \nonumber \\
				    &\times& f_r(E_{N \alpha}-E_{(N-1) \alpha'}), \\
\label{eq:outtunnel}
	W_{N \alpha, (N+1) \alpha'} &=& \sum_r \Gamma_{N \alpha, (N+1) \alpha'}^{r \sigma}  \nonumber \\
				    &\times& \left[ 1 - f_r(E_{(N+1) \alpha'} - E_{N \alpha}) \right],
\end{eqnarray}
where $f_r(E) = 1/(e^{(E - \mu_r)/k_B T} + 1)$ is the Fermi function of electrode $r$ with chemical potential $\mu_r$ and
$E_{N \alpha}$ is the energy of the molecular eigenstate $|N \alpha \rangle$.

Solving Eqs.~(\ref{eq:rateeq})--(\ref{eq:probnorm}) with the rates (\ref{eq:intunnel})--(\ref{eq:outtunnel}) gives 
the non-equilibrium occupation probabilities $P_{N \alpha}$ of all molecular many-body eigenstates. Based on these occupations, 
the current through the molecular junction can be calculated. The source--drain electron current can be obtained from
\begin{eqnarray}
\label{eq:current}
	I &=& \sum_{N \alpha} \sum_{N' \alpha'} W^I_{N \alpha, N' \alpha'} P_{N' \alpha'},
\end{eqnarray}
where the current rate matrix $\mathbf{W}^{I}$ is similar to (\ref{eq:intunnel})--(\ref{eq:outtunnel}), but including only tunnel
processes involving electrode $r = s$ and a plus/minus sign for electrons tunneling onto/out of the molecule. 

The needed input to the transport calculations is thus the full spectrum of molecular many-body eigenstates, 
$|N \alpha \rangle$, with energies $E_{N \alpha}$, as well as the hybridizations $\gamma^r_{i \sigma}$ between the single-particle 
orbitals $\psi_{i}$ and the lead electron states. 
The many-body states can be expressed as a sum of single-particle states 
(Slater determinants in first quantization language) with $N$ electrons:
\begin{eqnarray}
\label{eq:stateexpansion}
	|N \alpha \rangle = \sum_{\{ i_k \sigma_k\}} C^{\{ i_k \sigma_k\}}_{N \alpha} 
			    c_{i_1 \sigma_1}^\dagger \hdots c_{i_N \sigma_N}^\dagger |0\rangle,
\end{eqnarray}
where the sum runs over all collections $\{ i_k \sigma_k\}_{k = 1}^{N}$ satisfying $i_1 \le \hdots \le i_N$ 
and $|0\rangle$ is the state with zero valence electrons.
From the hybridizations $\gamma^r_{i \sigma}$, we can thus calculate the tunnel couplings by 
inserting~(\ref{eq:stateexpansion}) into~(\ref{eq:gammadef}).
However, finding the eigenstates (equivalent to finding all $C^{\{ i_k \sigma_k\}}_{N \alpha}$), 
in principle involves diagonalizing the many-body Hamiltonian 
of the molecule, excluding tunnel coupling to the electrodes, but \emph{including} the electrostatic environment of the electrodes, 
typically leading to reduced Coulomb repulsion and HOMO-LUMO gap due to screening by conduction electrons in the 
electrodes~\cite{Kaasbjerg08}. This is clearly a very difficult problem.  In the following we instead suggest a simple approximative approach, 
which still combines insights from ab initio quantum chemistry calculations with non-perturbative treatment of 
electron--electron interactions, crucial to correctly describe transport in the weak tunneling regime. 

The basis states $\psi_{i}$ are taken as the Kohn-Sham (KS) orbitals of the neutral molecule, calculated using DFT, 
and thus already include effects of electron--electron interactions on a mean-field level. 
We now assume that we can construct each many-body state 
by simply distributing the $N$ valence electrons over these orbitals, i.e., each state is given by setting one 
$C^{\{ i_k \sigma_k\}}_{N \alpha} = 1$ and all others to zero in (\ref{eq:stateexpansion}). 
This is justified when the MOs are not significantly changed upon charging, which is often the case. However, as long as 
the orbital asymmetries are not completely destroyed upon charging, the predicted NDR effect is not quantitatively affected.
(If, on the other hand, the change of the MOs as the molecule is charged is crucial, the above approach can 
simply be replaced by one where each state $|N \alpha \rangle$ is constructed instead from orbitals $\psi_{N i}$ 
calculated for the appropriately charged molecule.)
The "many-body" states are thus effectively single-particle states given by a single Slater determinant. 
The non-trivial effect of the electron--electron interaction instead enters in the corresponding 
energies $E_{N \alpha}$, which are approximated by the sum of the energies $\epsilon_{i \sigma}$ 
of the occupied orbitals (the collection of which we call $O_{N \alpha}$), plus a Coulomb charging term: 
\begin{eqnarray}
\label{eq:energy}
	E_{N \alpha} = \sum_{i \sigma \in O_{N \alpha}} \epsilon_{i \sigma} + 
		       \frac{1}{2} \sum_{(i\sigma, j \sigma') \in O_{N \alpha}} U_{ij},
\end{eqnarray}
where $U_{i j}$ is the Coulomb charging energy associated with one electron in orbital $i$ and one in orbital $j$.
$U_{i j}$ could be calculated within various approximations. 
However, the exact value depends on microscopic details of the junction~\cite{Kaasbjerg08} (not only the molecule),
which are at present uncontrollable, and unknown, in experimental situations. Fortunately, even though including a 
large Coulomb charging energy is crucial for a correct description of transport, the exact value typically 
has only a quantitative effect and can be seen as a fitting parameter when comparing with experimental data.

With the above approximations, the tunnel couplings~(\ref{eq:gammadef}) simplify to
$\Gamma_{N \alpha, N' \alpha'}^{r \sigma} = 2 \pi d_r |\gamma^r_{i \sigma}|^2$ if $|N \alpha \rangle$ is obtained from $|N' \alpha' \rangle$
by adding or removing one electron with spin-projection $\sigma$ in orbital $i$, and 
$\Gamma_{N \alpha, N' \alpha'}^{r \sigma} = 0$ if $|N \alpha \rangle$ 
and $|N' \alpha' \rangle$ do not differ only by the occupation of a single orbital.
The remaining problem is then to find the hybridization parameters $\gamma^r_{i \sigma}$.
Since tunnel amplitudes decrease exponentially with increasing barrier, it is reasonable to assume that 
tunneling only takes place into one or a few atoms at each end of the molecule which form the bond to the 
electrode. The natural basis is therefore the atomic orbitals (AOs), $\phi_j$. We express the MOs in terms 
of the AOs, $\psi_i = \sum_j a_j^i \phi_j$, which finally gives the tunnel coupling
\begin{eqnarray}
\label{eq:gammaresult}
	\Gamma_{N \alpha, N' \alpha'}^{r \sigma} = 2 \pi d_r \lvert \sum_j a_j^i \tilde{\gamma}_{j \sigma}^r \rvert^2,
\end{eqnarray}
were $\tilde{\gamma}_{j \sigma}^r$ is the hybridization of AO $j$ with electrode $r$ 
and $|N \alpha \rangle$ and $|N' \alpha' \rangle$ differ by an electron in orbital $i$. In principle, one could attempt to calculate
$\tilde{\gamma}_{j \sigma}^r$ using ab initio methods. However, the result would again strongly depend on the (unknown and uncontrollable)
details of the junction. In addition, in the weak coupling regime considered here, 
only the ratio of the source and drain tunnel couplings affect 
the transport behavior, while their overall magnitude only set the scale for current. The coefficients 
$a_j^i = \langle \psi_i | \phi_j \rangle$ can simply be read out of a DFT (or other quantum chemistry) program, provided 
that localized atomic orbitals are used as a basis. 
We note that if the KS orbitals would change substantially upon charging, such that we would be forced to use $N$-dependent orbitals,
$\psi_{N i}$,  orbitals in different redox states are not orthogonal and Eq.~(\ref{eq:gammaresult}) should be modified to contain a 
sum over all orbitals, weighted by the corresponding overlaps.
\bibliographystyle{apsrev}
\end{document}